\documentclass[a4paper,12pt]{article}
\usepackage{amsmath,amssymb,amsfonts,amsthm}

\usepackage[latin1]{inputenc}
\usepackage[english]{babel}
\usepackage{graphics}
\usepackage{amsfonts}

\usepackage[cmtip,all]{xy}
\usepackage[all]{xy}
\xyoption{all}
\xyoption{2cell}

\mathchardef\lt="313C 

\theoremstyle{definition}
\newtheorem{C*-category}[subsubsection]{Definition}
\newtheorem{example1}[subsubsection]{Example}
\newtheorem{example2}[subsubsection]{Example}
\newtheorem{E0}[subsubsection]{Definition}
\newtheorem{Fell bundle}[subsubsection]{Definition}

\newtheorem{example FB}[subsubsection]{Example}
\newtheorem{example 2 FB}[subsubsection]{Example}
\newtheorem{Fell bundle geometry}[subsubsection]{Definition}
\newtheorem{D}[subsubsection]{Definition}

\theoremstyle{plain}
\newtheorem{com 2}{Theorem}
\newtheorem*{allowed D}{Proposition 1}
\newtheorem*{generating set}{Proposition 2}

\newcommand{\C}{\mathcal{C}}
\newcommand{\Ob}{\textrm{Ob}}

\begin{document}

\title{Non-commutative fermion mass matrix and gravity}

\author{Rachel A.D. Martins \\\\ Centro de An\'alise Matem\'atica,\\ Geometria e Sistemas
Din\^amicas \\ Departamento de Matem\'atica \\ Instituto Superior T\'ecnico \\ Av. Rovisco Pais \\
1049-001 Lisboa \thanks{Email: rmartins@math.ist.utl.pt.}}

\date{\today}

\maketitle

\begin{abstract}
The first part is an introductory description of a small cross-section of the literature on algebraic methods in non-perturbative quantum gravity with a specific focus on viewing algebra as a laboratory in which to deepen understanding of the nature of geometry. This helps to set the context for the second part, in which we describe a new algebraic characterisation of the Dirac operator in noncommutative geometry and then use it in a calculation on the form of the fermion mass matrix. Assimilating and building on the various ideas described in the first part, the final part consists of an outline of a speculative perspective on (noncommutative) quantum spectral gravity. This is the second of a pair of papers so far on this project.
\end{abstract}

\\

\newpage

\tableofcontents

\newpage

\section{Introduction}

The main result is stated in the conclusions in section \ref{conclusions}. One of the mysteries of high energy physics is the form of the fermion mass matrix. Using a construction that unifies some ideas from non-commutative geometry and geometrical category theory, we carry out a calculation from which one may infer some information about the form of this matrix. The main point that we want to make is that the mathematical context we describe can be used to improve the understanding of the fermion mass matrix since the main principal is not model dependent, even though the model we are using is not up to date (even if we were to update it for example in the light of the recent right-handed neutrino data it may soon become obsolete again). The purpose of part I is to discuss and make connections among a small selection of the relatively recent developments in algebraic perspectives on canonical and covariant quantum gravity by various authors. We focus in particular on contributions involving ideas and 
methods from algebraic quantum field theory, category theory and spectral (noncommutative) geometry. One of the underlying themes of the paper will be that algebra can serve as a source of geometrical information, advocating using mathematics as a laboratory for physics rather than merely as a description of it. For example, category theory has proved to be an important resource for understanding quantum physics, the axiomatic approach to quantum field theory has been a success and methods such as index theory, the Tomita-Takesaki modular theory and Connes's spectral geometry can provide more powerful and general alternatives to the traditional differential geometry.

Even though it will probably at some point obtain a different consensual definition, just for the purposes of this paper we will use the term ``algebraic quantum gravity'' as a concise way of referring to the multifarious efforts towards the general aim of trying to construct background free and generally covariant theories of quantum gravity using concepts and machinery from category theory, AQFT and/or spectral triple geometry and especially attempts to reformulate all aspects of non-perturbative quantum gravity in algebraic language, including replacing diffeomorphims with algebra automorphisms and replacing the underlying point-set topology in Set with another category.

In the second part we develop the theoretical physics aspects of a project on algebraic quantum gravity for which the mathematical construction work was begun in \cite{sc}. It was not possible to produce a single longer paper on this project because mathematicians and physicists tend to have different expectations about what they wish to read. The main point is to suggest that Fell bundles might provide the mathematical ingredient needed to forge the link between AQFT, quantum gravity, the noncommutative standard model and category theory. In this two paper project we give a new algebraic characterisation of the Dirac operator for noncommutative geometry in order to begin to (i) converge the theory onto the experimentally determined fermion mass matrix, (ii) categorify the notion of real spectral triple and (iii) implement the tangent groupoid quantisation in the noncommutative standard model.

Our scope is confined to algebraic methods in loop and spin foam quantum gravity excluding string theories and conventional field theories generalised to noncommutative space-times by deforming the product of fields (sometimes called noncommutative quantum field theory) firstly because (a) the latter is already a well-established field with many people having already achieved important results in it, whereas this topic is new and the hope is to invite more people to study it, (b) the ultra-violet divergences problem is also dealt with (i) by the prediction of LQG that space-time be discrete on small scales\footnote{as a bonus, this even regulates the other interactions!} or (ii) by triangulating space-time, and there are other reasons for using noncommutative algebras in physics (not only to provide a fuzziness at the Planck scale), (c) there is the matter of personal choice as to whether one views it as axiomatic that a theory of gravity be background free and fully diffeomorphism invariant from the outset.

There are many reasons to involve algebra in quantum gravity either to reformulate it in that language or even just as a tool. One of the reasons for including part one as an introductory description was to elucidate some of them. For example, Connes suggests to extract physical information from mathematical entities in the following two senses. First he relates his classification of von Neumann algebras to a hypothesis about thermal time (with Rovelli) and secondly he develops a radical approach to modern physics (known as the noncommutative standard model, with Chamseddine): instead of following the convention of trying to quantise (the geometrical interaction) gravity, he \emph{geometrises} the `quantum' forces. This challenges the consensus that QED is already understood well enough, and raises the question as to whether unification should be worked out before quantisation instead of the other way around. While translating the geometrical theory of general relativity into algebraic language, the crucial 
step was to notice that the overall diffeomorphism group (which translates to the automorphism group) of the product of space-time with internal space is non-abelian and so the algebra of the configuration manifold pertaining to it is noncommutative. In addition, with the fermion mass matrix $M$ taking on a new mathematical identity as a Dirac operator, a new window opens:- to study $M$ is to study geometrical structures on noncommutative manifolds.

A second reason to involve algebraic methods in quantum gravity is to take seriously the following debate.
There is plenty of evidence to suggest that space-time is not actually best described by point-set topology.
This is one place in which modern physics and mathematics can diverge; physics is about trying to understand what is really there,
whereas when doing mathematics one feels obliged to take account of the fact that one can always think of a smaller or a bigger number. Some argue that mathematics as well a physics might benefit from a more physically meaningful approach to geometry.
 We describe below a little bit about how Crane uses concepts from Grothendieck categories and lessons from general relativity to propose the notion of a
quantum topos as an alternative to analytical mathematics for studying space-time. We infer - as he appears to suggest - that a fully algebraic theory of
quantum gravity will be based on a space-time that is fundamentally categorical in nature as opposed to having been triangulated and then categorified.

\subsection{Open questions: old and new}

In addition to suggesting new approaches to solving old problems in quantum gravity, the use of algebraic concepts and methods often does more to raise its own. Well known outstanding problems in the canonical and covariant approach - some of which will be discussed a little here - include the `issue of time' and its different interactions with the two approaches, defining and calculating observable quantities, making predictions about the physics of a singularity, the need for a choice of triangulation in spin foam models, verifying the gravitational classical limit of specific models, unification of the four forces, the fact of conflicting concepts with string theory, inclusion of fermionic matter into theories, unification with thermal physics, and resolving the debate as to whether space-time is by nature smooth or not.

And new open questions raised by algebraic quantum gravity include:

\begin{enumerate}
   \item What can we learn about nature from studying algebra?
   \item How can a background free generally covariant version of an AQFT be formulated?
   \item What might one mean by a non-perturbative theory of quantum gravity on the product of space-time with Kaluza-Klein internal space?
   \item Is there an uncertainty relation between charge and mass counterpart to those of Heisenberg?
   \item What is space-time like at scales towards the Planck length? For example, is it necessary to work outside point-set topology?
   \item Does curvature on a very small scale actually arise from space-time fuzziness or almost symmetry? (The fuzziness borrows length turning that into an amount of energy depending on $\hbar$ thus curving space while making mass-energy.)
   \item How can geometrical observables be defined over noncommutative spaces?
   \item How might the fermion masses be interpreted as quanta of internal space geometry in a precise way?
   \item What is the correct mathematically precise meaning of the interpretation of the Higgs field as a parallel transport between the chiralities?
   \item What is the right way to categorify noncommutative manifolds? And do we need higher categorical spectral triples?
   \item Can we define algebraic TQFTs and what algebraic invariants should they give rise to?
   \item How should the spectral path integral (see later) be discretised and viewed as a `partition function' of a spin foam model?
   \item How might the constraint equations be translated in terms of the spectrum of $D$? Will there be a single unified spectral constraint?
\end{enumerate}

Of course there are also specific technical problems, but these relate idiosyncratically to models and we will describe some of them.

\section{Part I: A review of algebraic perspectives on quantum gravity}

This is not supposed to be a literature review nor a summary of the content of each paper cited, rather its purpose is to extract and highlight certain contributions towards answering the old and new questions listed above. Each part is also interjected with comments to interconnect the topics and to sew central threads throughout the paper. All sincere apologies to anyone whose work has been omitted or misrepresented.

\subsection{Spectral geometry}

Assuming some familiarity with the noncommutative standard model \cite{sap} \cite{gravity}  by Connes and Chamseddine (see also \cite{ncg and sm} and \cite{forces} by Sch\"ucker)  we begin by briefly reviewing it as an algebraic formulation of general relativity and explaining how its defining algebraic concepts were used to learn something new about the relationship between classical gravity and the standard model. Then we focus on the new point of view it led Connes to on the path integral formulation of quantum gravity (\cite{essay}).

The noncommutative standard model is of course the model of gravity on the total space consisting of a 4-dimensional space-time manifold (all calculations are usually restricted to the Euclidean signature) component and a finite noncommutative ``internal'' Kaluza-Klein type space where the gauge and scalar sectors of the standard model are the internal space component of the gravitational action and in which the standard model spectral triple Dirac operator plays the role of the gravitational field. It is a fully background free, generally covariant extension of general relativity formulated in pure algebraic terms. Since the Dirac operator on internal space is the fermion mass matrix, the problem of quantisation makes the latter another mystery of quantum gravity.

The treatment of diffeomorphisms in this noncommutative context is the following. First recall that a diffeomorphism acting on a manifold $M$ is translated into algebraic language as an automorphism of the algebra $C(M)$. The algebra of the product space is

\begin{equation}
 A = C^{\infty}(M) \otimes A_F
\end{equation}

where $A_F$ is a finite dimensional semi-simple algebra whose role is to capture the charges and the chiralities of the `internal space' of the standard model or a variation of it. Gauge transformations are implemented by inner automorphisms and hence are treated on the same footing as diffeomorphisms. Connes's algebraic formulation of the equivalence principle is to `fluctuate' the Dirac operator: if $D$ is the flat Dirac operator on the total product space and $D_F$ is its internal space component,

\begin{equation}
  D = \gamma^{\mu} \partial_{\mu}  + \gamma^5 \otimes D_F
\end{equation}

the fluctuated Dirac operator is given by a set of linear combinations of automorphisms of the total space algebra $\mathcal{A} = A \otimes A^{opp}$,

\begin{equation} \label{fluctuations}
  D^f = \sum_{\textrm{finite}} r_j L (\sigma_j) D L(\sigma_j)^{-1}, ~~r_j \in \mathbb{R}, ~~ \sigma_j \in \textrm{Aut}(\mathcal{A})
\end{equation}

giving rise to a general curved Dirac operator on the total space, and the configuration space of the theory is the set of all possible such $D^f$. The spectral action is a function of the eigenvalues of $D^f$ and since it only depends on them, it is diffeomorphism invariant. The action is that of gravity on the total space and as a result its calculation results not only in the Einstein-Hilbert action but also the  scalar sector plus the gauge sector come out. Thus we find our first example of a calculation in which algebra has been used as a laboratoy for studying physics.

Before commencing a quantisation program for spectral gravity one might hope for an algebraic formulation of true general relativity to be already in place. However, starting with an initial Dirac operator and then fluctuating it is similar to perturbation theory, not entirely reflecting the principle we learned from Einstein that gravity is geometry and therefore the gravitational field variable ($D$) should encode all possible geometries. This led Barrett \cite{smv} to suggest calculating internal space Einstein's equations such that the configuration space consist of all possible Dirac operators $D_F$ satisfying Connes's definition of it. That study resulted in the question: Why is it that the fermion mass matrix of the standard model is not a solution to those equations? Now that its context has been explained, we can add this to the list of open questions in algebraic quantum gravity.

Now we focus on the paper \cite{essay} by Connes. An important conceptual point he makes is that the spectral point of view allows one to take into account the lessons from both quantum field theory and general relativity, by which he means for example that the spectral action is background free and tries to understand space-time on a very small scale. He goes on to remind us that a straightforward application of QFT to the gravitational field does not work and that instead he will write down a sum of geometries as a spectral path integral. Next there is a fresh introduction to renormalisation. The Hopf algebra structure behind the Feynman diagrams was worked out with Kreimer offering a new insight into calculations in physics that are not fully understood mathematically.

Connes goes on to paint a vivid picture of what he sees in the phrase ``NCG spectral paradigm'' in contrast to the ``Riemannian $g_{\mu  \nu}$ paradigm''. The first point is that the conventional wisdom on the total group of symmetries in physics is

\begin{equation}
 \mathcal{U} = (M, \mathrm{U}(1) \times \mathrm{SU}(2) \times \mathrm{SU}(3)) \ltimes \mathrm{Diff}(M)
\end{equation}

and therefore the space one should be looking for for this to be the symmetries of, is not a convential manifold but a noncommutative space.

The second point is a very deep philosophical one. It cannot be put in a nutshell, but he suggests that our paradigm of geometry should be adapted to the standard of length we see in nature and that comes from spectral physics.

Now we come to quantum spectral gravity. To write down a sum over geometries one needs to know how to classify 4-manifolds and one has to choose a way of triangulating them. However, in the paradigm of spectral geometry, it is only the space of Dirac operators that is required. Connes's ``first shot'' at writing down a functional integral for quantum spectral gravity,

\begin{equation}   \label{spectral path integral}
 < \mathcal{O} > = N \int \mathcal{O} e^{-\mathrm{Tr}(f(D)) - <\bar{\Psi},D \Psi> - \gamma^5(c,D)} D[\Psi]D[\bar{\Psi}]D[D]D[c]
\end{equation}

where $\mathrm{Tr}(f(D))$ is the spectral action, $<\bar{\Psi},D \Psi>$ is the fermion sector of the standard model, $N$ is a normalisation factor and $c$ is the volume form.

\subsection{Spin network algebras}

In \cite{Perez Rovelli} Perez and Rovelli provide a connection between Algebraic quantum field theory and Quantum gravity using group field theory methods and the GNS construction.

Normally, field theories begin with a field configuration space and one defines an algebra and a Hilbert space in its terms, in other words the geometrical aspects of the theory come first and then one builds on that. One of the central ideas of Wightman's axiomatisation of quantum field theory was that one could begin with an algebra and reconstruct the theory from it. However, it was still necessary to make a choice of Poinar\'e group representation for the dynamics but Haag and Kastler's work on algebraic quantum field theory (see \cite{Haag's book}) together with some ideas of Connes and Rovelli (see below) shows that the algebra can fully capture the entire theory including the dynamics. Hitherto, AQFTs describe fields only relative to a background manifold and not geometrical degrees of freedom. In \cite{Perez Rovelli} Perez and Rovelli take a step towards algebraicising quantum gravity leading towards a background free AQFT.

Perez and Rovelli define the algebra $A$ to be that of the functions on a space of (diffeomorphism equivalence classes of) spin network states $s$ and then they reconstruct the Hilbert space via the GNS construction for $A$. The observable quantities are called ``$W$-functions'' (W for Wightman), coming from linear functionals over $A$, specifically, transitions functions of a field $\phi_s$ over a group and can be thought of as probability amplitudes for transitions between 3-geometries (the spin network states),

\begin{equation}
 W(s) = \int [D \phi] \phi_s e^{iS[\phi]}
\end{equation}

where $\phi_s$ is a field defined over a group manifold (as in group field theory where it is built out of the group representation theory instead of starting with a spin network and labelling it with representations and intertwiners of the group). The $W$-functions are gauge invariant and diffeomorphism invariant and so are physically interpretable quantities. Enough of the technical detail is worked to enable the authors to present some specific models.

Group field theory or GFT might be important in  algebraic quantum gravity because the fundamental object is a group with its representation theory rather than a space of connections or a space-time manifold that you have to triangulate and then label with the group representations. The aim of GFT is to produce a background independent, non-perturbative description of the dynamics of space-time in the context of simplicial gravity \cite{GFT}. It is more than a new tool for quantum gravity as it is an enveloping theory for the canonical and covariant approaches, combining and superceding them. Its proponents state that ``Spin foams are nothing other than Feynman graphs of a GFT'' and can show that it is enough to reproduce any result obtained in the spin foam context.

In the next paper we review, the algebraicisation of gravity is taken further from the conceptual point of view in that it discusses a fully general covariant approach and a deep mathematical reason why one can by-pass the need for a choice of Poincar\'e representation or Hamiltonian.

\subsection{The thermal time hypothesis}

In the paper \cite{thermal} one finds what may be the most important example of Algebra being used as a resource in the study of Theoretical physics. To address the issue of time in quantum gravity Connes and Rovelli build on the relationship between KMS theory and Tomita-Takesaki theory, which Haag \cite{Haag's book} celebrates as a `pre-stabilised harmony' between physics and mathematics. The reader will find informative expositions of modular theory and its relation to quantum systems and KMS theory in \cite{thermal} and in \cite{Haag's book}.

The authors begin by reminding us that a quantum system that is not generally covariant consists of observables and states for the kinematics and for the dynamics a specific way for the states to evolve in time encoded in a choice of Poincar\'e represenation or equivalently a Hamiltonian, whereas generally covariant theories are fully diffeomorphism invariant (such as general relativity) and so there cannot be a singling out of a direction to be labelled as `time'. For this reason, generally covariant theories have this problem of time, for example in spin foam models there is no canonical emergence of a physical time evolution. Interestingly, they go on to observe that the emergent notion of time in general relativity is state dependent. And they also point out that not only is general relativity incompatible with quantum mechanics but also with thermal dynamics for example because the Gibbs state determines an overall time-flow for the system.  Their `radical' solution is to formulate a \emph{state 
dependent notion of time} (where the state is a positive linear functional over an algebra). Recall that any faithful state $\omega$ is KMS (thermal) with respect to the modular automorphism group $\sigma^{\omega}_t$ it itself generates. Their \emph{thermal time hypothesis} states that,

``The physical time depends on the state. When the system is in a state $\omega$, the physical time is given by the modular group $\sigma_t$ of $\omega$.''

In multi-particle sytems one expects there to emerge an overall time-flow as a state independent phenomenon; the Radon-Nikodym theorem states that all modular groups of a given von Neumann algebra are equivalent in the sense of being inner automorphic. This exposes the deep physical nature of a von Neumann algebra and as it is expressed in \cite{thermal}, ``\emph{A von Neumann algebra is intrinsically a dynamical object.}''

In \cite{Pierre} Martinetti reviews and makes connections between the thermal time hypothesis and spectral triple physics, and discusses a physical test of the hypothesis in the Unruh effect. He suggests that the Planck scale discreteness (as predicted by loop quantum gravity) could be due to a phenomenon resulting from what may be the noncommutative nature of space-time and explains why the degree of fuzziness or scale of noncommutativity might soon be measurable experimentally.

\subsection{Categorical physics and quantum topoi}

In his recent work \cite{qg},  \cite{quantaloids}, \cite{causal sites}  Crane proposes that the main obstruction to progress in quantum gravity at the moment is due to the fact that it inherits issues from general relativity so that classical limit calculations are flawed, the problem is that Riemannian geometry is based in traditional analytical methods, where it is assumed that it makes sense to consider infinitely small distances but this is not physically meaninful. Of course general relativity has been used to do calculations\footnote{GPS!} with excellent precision but the continuum approximation is not enough to do quantum gravity throughout its greater realm of validity. Crane reminds us that two of the key reasons are that the continuum makes no sense whatsoever below the Planck scale and that a bounded region can only contain finite information. Ironically then, what is needed is mathematics that is more sophisticated than analysis but where the Hilbert spaces are finite dimensional. This implies 
that the main reason why the Barrett-Crane model \cite{BC model} hasn't been fully accepted as a true candidate for a theory of quantum gravity - because it doesn't have a classical limit in Riemannian geometry yet - is therefore not really valid.

So what should the continuum mathematics be replaced with? Einstein allowed his physical intuition to lead him to Riemannian geometry, which was the correct answer for general relativity. Crane suggests doing the same to find the right mathematics to address the challenges that gravity faces now and that this will need to involve (among others) the following features: a non-distributive lattice of observable regions; space-time with an intrinsically categorical description; be able to describe the finite flow of information between regions; points should appear only relative to the observer and there should be a minimum length-scale; noncommutative algebras. In this ``relational setting'' the classical limit will be a formulation of correlations between where different observers see the same probe. The mathematics this leads to is a ``causal site'' \cite{causal sites} (with Christensen),  taking inspiration from causal sets and Grothendieck topologies. A \emph{causal site} is  set of ``regions'' with 2 
binary relations and satisfying 9 axioms. A related notion for a description of quantum space-time is that of a \textit{quantum topos}, (\cite{quantaloids}) which is the category of sheaves over a quantaloid. An ordinary topos is a distributive lattice (does not allow for quantum logic) but a quantum topos does not have that defect. Moreover, any topos can be reconstructed from the sheaves over some quantaloid.

The relational quantum space-time is similar to the categorical structures underlying algebraic quantum field theory. In the latter one has a functor $ \mathcal{O} \mapsto A(\mathcal{O})$ from regions of (typically Minkowski) space-time into a net of observable von Neumann algebras, whereas in the former $R \mapsto Q(R)$ regions of curved space-time are mapped functorially into a ``$Q$-structure'' such as a quantaloidal sheaf. However, $R$ can be an object in a 2-category, the space-time can be curved and the whole system must be background free - to achieve this, one considers whole categories of $Q$-structures so that the space-time is not fixed. In Crane's point of view there is much more emphasis on topics in category theory, such as topos theory, homotopy theory, model categories, quantaloids, Grothendieck sites, higher category theory and applying these to physics such as in categorical state sum models.

Category theory is not just a methodology for collecting, sorting and classifying mathematical objects (even in pure mathematics) but it has a deep geometrical significance; categorical physics \cite{category qg,clock} can provide an algebraic characterisation of space-time and unifying quantum mechanics with relativity might require a lot of it. The Riemannian geometrical nature of classical space-time  leads to the equivalence principle that laws should be expressed in tensor form and the \emph{categorical nature of space-time} implies the principle in quantum gravity that laws must be expressed in \textit{functorial} form (\cite{causal sites}).

\subsection{Spectral triple and $C^*$-categories}


In a collection of articles including \cite{Paolo2}, \cite{Paolo}, \cite{BCL survey 2010} Bertozzini, Conti and Lewkeeratiyutkul construct a categorification scheme for spectral triples drawing motivation from categorical physics, and because,  ``Although the main strength of noncommutative geometry is a full treatment of noncommutative algebras  as `duals of geometric spaces', the foundation relies on...'' category equivalences and anti-equivalances such as the Serre-Swan theorem, Gelfand-Naimark theorem and they give several more examples. So, if we are to view real spectral triples and their reconstruction theorem as a geometrical counterpart to the Gelfand-Naimark theorem, then we should define morphisms between triples: indeed the authors work through a counterpart of the Gelfand-Naimark theorem, categorifying the reconstruction theorem for Riemannian spin manifolds \cite{BCL Hoz}, plus many more results about equivalences and anti-equivalences between the new categories they define. These authors 
interpret ``spectra'' for $C^*$-categories to be what they call `topological spaceoids' (see for example \cite{Paolo}, \cite{BCL Hoz}), to construct categorified Gelfand theory.

Here is an introductory example of the kind of category defined in the work cited above. A morphism in the category of spectral triples consists of a $*$-homomorphism $\phi$ between the algebras and a bounded linear map $\Phi$ intertwining representations of the two algebras on the two Hilbert spaces. They also define two subcategories, if the objects are real spectral triples then $\Phi$ must satisfy the additional condition that $J_2 \circ \Phi = \Phi \circ J_1$ where of course $J_1$ and $J_2$ are the reality operators belonging to each of the two real structures, and a morphism of even spectral triples satisfies
$\chi_2 \circ \Phi = \Phi \circ \chi_1 $.

They also define an isometry category of spectral triples: a ``metric morphism'' of compact spectral triples,

\begin{equation}
 \phi: (A_1, H_1, D_1) \rightarrow (A_2, H_2, D_2)
\end{equation}

is a unital epimorphism $\phi: A_1 \rightarrow A_2$ whose pull-back $\phi^*: P(A_2) \rightarrow P(A_1)$ is an isometry, that is,

\begin{equation}
       d_{D_1}(\phi^*(\omega_1), \phi^*(\omega_2))    =      d_{D_2}(\omega_1, \omega_2) ~~ \forall \omega_1, \omega_2 \in P(A_2)
\end{equation}

noticing that since $\phi$ is an epimorphism, its pull-back maps pure states $P$ into pure states.

(Work on categories of spectral triples has also been carried out by Mesland \cite{MarcolliCat,Mesland 2011}).

\subsection{Spectral triples in Loop quantum gravity}

Aastrup, Grimstrup and Nest's program of combining noncommutative geometry with quantum gravity now consists of several papers of which some examples are \cite{AG1}, \cite{AG2}, \cite{GAN2}. Unfortunately this subsection is not representative of the current state of the art but we give a brief introduction. For quantum gravity reviews see \cite{Baez review}, \cite{Perez review}, \cite{Rovelli review}.

Of course the details vary from model to model but the basic proforma for constructing loop quantum gravity models is more or less the following. One begins at the kinematical level by setting up the algebraic and geometrical structures for the theory in order to define a $C^*$-algebra of observables $A$ (as functionals of the field configuration variables), a set of measurable quantities and a kinematical Hilbert space. The geometrical structure is a choice of $G$-bundle over a nested sequence of graphs and the observable algebra is generated by the traces of the holonomies (closed loop parallel transports): functionals on the projective-limit manifold or pro-manifold consisting of the space of all connections $\mathcal{A}$  on the inductive system of graphs, which gives the configuration space of the theory and together with the momentum conjugate variable, the driebeins $E$, the quantum Poisson algebra is constructed to give a bounded and analogous version of the famous $[p,q] =  i \hbar$ commutation 
relation. The kinematical Hilbert space is $\mathcal{L}^2(\mathcal{A})$ but this must be reduced to a set of physical states. The physical Hilbert space $H_{\mathrm{phys}}$ is built from an orthonormal set of diffeomorphism classes of spin network states (modulo the other constraints) and it must carry a representation of $A$ and of the quantum Poisson algebra. The observable quantities are given for example by evaluating linear functionals over $A$. Finally, the dynamics of a quantum system should describe the physical evolution of the states and involves calculating the observable quantities by solving the Hamiltonian constraint or by adopting the spin foam cousin approach of discretised path integral methods.

Aastrup, Grimstrup and Nest construct a reformulation of loop quantum gravity in which the machinery of spectral triple noncommutative geometry is employed to study geometrical structures over a space of connections. This is a non-perturbative, background independent quantum field theory of gravity with most of the physical intuition coming from loop quantum gravity and with mathematical methods from spectral triple geometry. Even though they use the word ``machinery'' to refer to the latter, in many ways they carry the concepts of spectral triples beyond the descriptive tool sense.

What they gain from their approach includes new techniques for studying spaces of connections in loop quantum gravity, a new algebraic interpretation of the latter and also a step towards quantising the noncommutative standard model.

The authors construct a Dirac operator on the above ``pro-manifold'' by assigning a Dirac operator $D_i$ to the space of connections on each graph $\Gamma_i$ in the nested sequence. Instead of considering all graphs, they restrict to a system of triangulations (or in recent work a cubic system was adopted) defining a limiting triple

\begin{equation}
 (B_{\Delta}, D_{\Delta} = \sum a_i D_i, H_{\Delta})
\end{equation}

where $B_{\Delta}$ is an algebra whose spectrum is identified with the pro-manifold of connections $\mathcal{A}$, precisely defined as the $C^*$-completion of the algebra generated by the holonomies of the connections in $\mathcal{A}$. The $\{ a_i \}$ form an infinite sequence carrying metric data. They show that when its limit approaches infinity fast enough then $D$ has compact resolvent and then the spectral triple $(B_{\Delta}, D_{\Delta}, H_{\Delta})$ is a (semi-finite) spectral triple. Due to the triangulation aspect of the construction, the Hilbert space is invariant only of a discrete diffeomorphism group but this is the only essential difference with the physical Hilbert space of loop quantum gravity.

They show that the interaction between $B_{\Delta}$ and $D_{\Delta}$ reproduces the Poisson structure of general relativity and also relate the square of the flux operator with the Dirac operator.

Note that as no foliation is required to construct the model and that there is no need to make a choice of dimension, it is generally covariant in all respects from the outset.

Differences with the traditional loop quantum gravity include the crucial one that the holonomies generating the algebra are left untraced, which results of course in a noncommutative algebra, it is fully general covariant from all points of view, there are more tools in the box and more powerful ones coming from the algebraic flavour, it allows for the virtual ``fifth dimension'' for internal space and there is a partial solution to the diffeomorphism constraint due to the triangulation approach.

Open problems they discuss include involving the thermal time hypothesis (see review above), to deepen understanding of the sequence of $a_is$, how to reproduce more of the results of loop  quantum gravity, how to reproduce the action of the noncommutative standard model and adding fermionic matter - the fermions are not the same as those of the standard model because they live on $\mathcal{A}$ rather than space-time and they suggest that to reproduce the classical spectral action they would need some kind of $\delta$-function on $\mathcal{A}$ to pick out the one to leave only the space-time degrees of freedom for the fermions. They hope that in their formulation of quantum gravity matter degrees of freedom will emerge as an automatic consequence of their construction.

\subsection{q-deformed symmetries}

Unfortunately this subsection will be very short due to time constraints. There is a series of 3 papers by Majid including \cite{Majid} in which he discusses his algebraic approach to quantum gravity in which quantum groups describe symmetries in physics. He suggests that classical limits of quantum gravity models should be calculated in the context that the space-time be noncommutative instead of the usual attempts to find a limit in the continuum mathematics that Riemannian geometry is based on. This reflects Crane's argument (see above) that the classical limit should be pursued in terms of a refinement of general relativity taking the non-continuum nature of space-time into account. Majid studies Poincar\'e Hopf algebras as a particularly important example of his decription of almost symmetries. He asks and addresses the following two questions:

\begin{enumerate}
 \item How could we see a noncommutative plane wave?
  \item   If we smash together two waves of non-abelian momenta $p$ and $p'$, which way around do we form the composite?
\end{enumerate}

\section{Part II: Fell bundle quantum systems}

The purpose of this part is to develop some of the physical ramifications of a project in algebraic quantum gravity in which the foundational mathematical construction work was begun in \cite{sc}. In that paper a categorification program for real spectral triples was worked through using ``Fell bundle geometries'' as a first step towards algebraicising quantum gravity (a list of motivations for categorifying real spectral triples appears in the introduction to \cite{sc} which we won't repeat here, except to refer to 2.4 above and the idea that space-time or internal space is intrinsically categorical). Some of the the physical aspects were alluded to there but here we treat them more thoroughly.

Below we explain how to build a Fell bundle geometry and the new algebraic characterisation of the Dirac operator, or ``Fell bundle Dirac operator'' and then show that the finite spectral triple for one generation of fermions has a categorification in this way. We learn from the noncommutative standard model that to study the fermion mass matrix is to study differential structures on noncommutative manifolds and we find that the interpretation of the above said spectral triple as a Fell bundle geometry results in a fermion mass matrix with a form closer to that given by experiment.

Secondly we use the $C^*$-algebraic structure of Fell bundles to aim to put in place some of the initial building blocks for a candidate non-perturbative (background free) generally covariant quantisation of spectral gravity. The basic variable to be quantised,   instead of the connection or the holonomy, will be the Dirac operator. We can use concepts from loop quantum gravity and spin foams as guiding principles but many of the features will be different for that reason and also of course because the space-time is noncommutative and for that reason we pass to categorical methods beyond those of Set to deal with the ``spectrum'' of the algebra.

\subsection{Preliminaries}

In this subsection some less well-known definitions are recalled but time-constrained physicists are invited to skip directly to the examples. Topics in theoretical physics such as algebraic quantum field theory (or AQFT), category theory, non-perturbative quantum gravity and spectral triple physics are omitted here.

\subsubsection{$C^*$-bundles and Fell bundles}

We begin with a definition from Dixmier's book \cite{Dixmier}. To avoid overloading this section we
will not also give the definition of a continuous field of Banach spaces, which can also be found
in Dixmier. In this paper we refer to a continuous field of $C^*$-algebras as a $C^*$-bundle and denote it $E^0$ and the $C^*$ completion of its algebra of sections $C^*(E^0)$.

\begin{E0}[continuous fields of $C^*$-algebras] Let $\mathcal{A} = (A(t), \Theta)$ be a continuous
field of Banach spaces or Banach bundle over $M \ni t$, denoting the sections by $\Theta$ and each
$A(t)$ a $C^*$-algebra. Then $\mathcal{A}$ is a continuous field of $C^*$-algebras if and only if
there exists a total subset $\Lambda$ of $\Theta$ which is closed under multiplication and
involution. Let $A$ be the set of sections $x$ whose length is 0 on the Banach space over
$t=\infty$. Then put on $A$ the norm: $\vert \vert x \vert \vert = \mathrm{sup}_{t \in T}~ \vert
\vert x (t) \vert \vert \lt~ +~ \infty$. This makes $A$ a $C^*$-algebra which is called the
$C^*$-algebra defined by $\mathcal{A}$.
\end{E0}

\begin{Fell bundle}[Fell bundle]   \label{defining list}

A Banach bundle over a groupoid $p:E\rightarrow \Gamma$ is said to be a Fell bundle
\cite{fbg} if  there is a continuous multiplication $E^2 \rightarrow E$, where

\begin{displaymath}
 E^2 = \{(e_1,e_2)~ \in~ E \times E ~|~ (p(e_1),p(e_2))~ \in~ \Gamma^2\},
\end{displaymath}

and an involution $e \mapsto e^{\ast}$ which satisfy the following axioms ($E_{\gamma}$ is the fibre $p^{-1}(\gamma)$).

\begin{enumerate}
 \item $p(e_1e_2) = p(e_1)p(e_2)~~ \forall~ (e_1,e_2) ~\in ~E^2$;
 \item The induced map $E_{\gamma_1} \times E_{\gamma_2} \rightarrow  E_{\gamma_1 \gamma_2}$,~~ $(e_1,e_2) \mapsto
e_1e_2$ is bilinear $\forall ~(\gamma_1, \gamma_2)~\in~\Gamma^2$;  \item $(e_1e_2)e_3=e_1(e_2e_3)$ whenever the
multiplication is defined;
 \item $\Vert e_1e_2 \parallel~~ \leq ~~\parallel e_1 \parallel ~ \parallel e_2 \parallel ~~\forall~(e_1,e_2)~\in~E^2$;
 \item $p(e^{\ast})=p(e)^{\ast}~~\forall~e~\in~E$;
 \item The induced map $E_{\gamma} \rightarrow E_{\gamma^{\ast}},~~e \mapsto e^{\ast}$ is conjugate linear for all $\gamma ~\in~\Gamma$;
 \item $e^{\ast \ast} = e ~~\forall~ e ~ \in ~ E$;
 \item $(e_1e_2)^{\ast} = e_2^{\ast} e_1^{\ast} ~~\forall~(e_1,e_2) ~\in~E^2$;
 \item $\parallel e^{\ast} e\parallel ~=~ \parallel e \parallel ^2~~\forall~e ~\in ~ E$;
 \item $e^{\ast} e \geq 0 ~~\forall~ e ~ \in ~ E$.
\end{enumerate}

\end{Fell bundle}

\begin{C*-category}\cite{GLR},\cite{Mitchener}
A \emph{$C^*$-category} is a category $\C$ in which for all objects $A,B\in\Ob_\C$, the homsets $\C_{AB} := \mathrm{Hom}_{\C}(B,A)$ are complex Banach spaces, the compositions are bilinear maps such that $\parallel xy \parallel ~ \leq ~ \parallel x \parallel \cdot \parallel y \parallel$ for all $x \in \C_{AB}$, $y \in \C_{BC}$ and there is an involutive antilinear contravariant functor $* : \C \rightarrow \C$ preserving objects such that $\parallel x^*x \parallel ~ = ~ \parallel x \parallel^2$ and such that $x^*x$ is a positive element of the $C^*$-algebra $\C_{AA}$ for every $x \in \C_{BA}$ (i.e. $x^*x = y^*y$ for some $y \in \C_{BA}$).
\end{C*-category}

Note that each $\C_{AA}$ is a $C^*$-algebra and also that it possesses a unit element due to the identity axiom in category theory. (Reference to category theory: \cite{cwm}.)

\begin{example1}\cite{GLR}
The category of Hilbert spaces and bounded linear maps.
\end{example1}

\begin{example2}\cite{GLR}
The category Rep($A$) of representations of a $C^*$-algebra $A$ on a Hilbert space and intertwining operators.
\end{example2}

Recall that if a groupoid is \textit{principal}, it means that there is at most one pair of arrows
between any two units and this is the same thing as an equivalence relation, which is a
subgroupoid of a pair groupoid or maximal equivalence relation. A unital saturated Fell bundle over a maximal equivalence relation or pair groupoid is equivalent to a full C*-category, which is a category of Morita equivalence bimodules over C*-algebras. In \cite{sc} we referred to these categories as ``Fell bundle C*-categories''.


To see that the above is a $C^*$-category, one views the elements of the Morita
equivalence bimodules as morphisms and the $C^*$- algebras over the groupoid
units as objects. These have to be unital algebras to satisfy the category theoretic unit law and
finally the multiplication in the Fell bundle is associative by definition. This is a
small subcategory of the category of Banach spaces together with a *-functor $\pi : E \rightarrow
G$. The practical implication of $E$ being \textit{saturated} ($E_{\gamma_1}.E_{\gamma_2}$ is total
in $E_{\gamma_1 \gamma_2}$ for all $(\gamma_1, \gamma_2)~\in~\Gamma^2$) is that the latter are
Morita equivalence bimodules. Note that finite dimensional Fell bundles are saturated. As is true
for any $C^*$-category, it can be represented on a concrete $C^*$-category, (that is, a small
subcategory of Hilb\footnote{objects are Hilbert spaces and morphisms are the bounded linear maps
between them.}). See \cite{GLR}, \cite{Paolo}.

The following two examples of Fell bundles are $C^*$-categories:

\begin{example FB} Consider the
pair groupoid $\Gamma$ on two objects. The fibre over each object is a C*-algebra:  $E_{\gamma
\gamma^{\ast}}$ and $E_{\gamma^{\ast}\gamma}$, and the fibre over each of the remaining two arrows
$\gamma$ and $\gamma^{\ast}$ is a Banach space, $E_{\gamma}$ and $E_{\gamma^{\ast}}$. These two
Banach spaces are modules over the said two $C^{\ast}$-algebras. This Fell bundle is saturated and
the Morita equivalence linking algebra is given by:

\begin{equation} \label{linking algebra}
\left(   \begin{array}{cc}
  E_{\gamma \gamma^{\ast}}       &    E_{\gamma}     \\
   E_{\gamma^{\ast}}      &       E_{\gamma^{\ast} \gamma }
\end{array}  \right)
\end{equation}

which is the algebra of sections of this Fell bundle.

\end{example FB}

\begin{example 2 FB} Let $E$ be a complex line Fell bundle over an r-discrete groupoid. The algebra
of sections of $E$ is isomorphic to the groupoid algebra. For example, the sectional algebra of the
line bundle over the groupoid $G=\mathrm{Pair}(M)$ is $M_n(\mathbb{C})$ where $n$ is the number of
units in $G$ or points in $M$, which we identify with $G_0$. $M_n(\mathbb{C})$ is the linking
algebra of the Morita equivalence bimodules which are the morphisms in the category $E$. The $C^*$-bundle algebra $C^*(E^0)$ is $\bigoplus_n \mathbb{C}$.

It is easy to notice that any algebra $B$ is Morita equivalent to itself with $B$ as the imprimitivity bimodule. For example, the set of homomorphisms between two fibres of $E^0$ or Hom$(\mathbb{C},\mathbb{C})$ is a Morita equivalence (or imprimitivity) $\mathbb{C}$ - $\mathbb{C}$ bimodule and  is of course isomorphic to $\mathbb{C}$.
                     \end{example 2 FB}

\subsubsection{The tangent groupoid quantisation}

We give  a very brief overview of the idea of the tangent groupoid. For more details we cite
\cite{Connes's book}. It is a quantisation procedure through asymptotic morphisms. For a particle
system on a Riemannian manifold $M$ (or on the prototypical case of $\mathbb{R}^n$, or even on a
more difficult space) the cotangent space captures the phase space and its $C^*$-algebra is taken
to be the algebra of observables. This is commutative, so deformation quantisation methods are
needed. In the tangent groupoid the Moyal deformation is used. The tangent groupoid is:-

\begin{displaymath}
 \mathcal{G}M = TM \times \{  0 \}  \cup  M \times M \times (0,1]
\end{displaymath}

where $TM$  is a groupoid, $M \times M$ is the pair groupoid (equivalence relation), and
$\mathcal{G}M$ is therefore itself a groupoid. Instead of $\hbar$ always taking a certain value, it
is viewed as a continuous parameter taking values in an interval of the real line $[0,1]$ and the
classical limit is obtained as it `goes to zero'. The $C^*$-algebra of the tangent groupoid is the
union of a continuous\footnote{that is, norm continuous} field of $C^*$-algebras $A_{\hbar}$ over
the set of $\hbar$s. The asymptotic morphism is a morphism from the algebra $A_0$ over $\hbar = 0$
to any of those over $\hbar \neq 0$. The algebras over $\hbar \neq 0$ are noncommutative, they are
the algebra of the pair groupoid, which is the compact operators on the Hilbert space $L^2(M)$. To
emphasize the role of the groupoid elements as its generators, we will denote this algebra as
$C^*(M \times M)$.



\subsection{How to build a noncommutative manifold from a Fell bundle}

Below we clarify the conceptual steps involved in the construction of a noncommutative Fell bundle geometry $(E,D)$, the mathematical development of which was begun in \cite{sc}. In the subsection to follow we also give an alternative more general and more transparent definition of a Fell bundle Dirac operator.

Let $E$ be a Fell bundle $C^*$-category over a groupoid over a compact space $M$. If $M$ is smooth, let it be simply connected. Since their base spaces are groupoids, viewing Fell bundles as noncommutative geometries naturally leads to a quantisation scheme for a noncommutative space-time via the tangent groupoid quantisation. That is, they provide a generalisation of the tangent groupoid quantisation to systems in which the underlying space-time (or internal space) is noncommutative. As the tangent bundle $TM$ is deformed to the pair groupoid $\Gamma = M \times M$, one replaces the commutative algebra $C^*(TM)$ with the noncommutative groupoid algebra $C^*(M \times M)$. For $E$ a complex Fell line bundle over $(\Gamma)$, we have $C^*(E) = C^*(M \times M)$ and $C^*(E^0) = C^*(M)$. This allows us to make the generalisation to virtual  when $E$ is a more general Fell bundle over $\Gamma$: let $E$ be a Fell bundle $C^*$-category over a discrete compact pair groupoid $\Gamma$ (for example its fibres may be $M_n(\
mathbb{C}$ instead of $\mathbb{C}$ and each fibre has a virtual spectrum instead of a point in $M$). Then the configuration algebra is $C^*(E^0)$ and the coordinate algebra is $C^*(E)$ and they are \textit{both} noncommutative. The Hilbert space $H$ is finite dimensional and carries a faithful representation of both algebras.

To complete the construction of a Fell bundle geometry we assign to it a Fell bundle Dirac operator, see below.

\subsection{A new algebraic characterisation of the Dirac operator}

This part is to explain why the Fell bundle Dirac operator was defined in the way it was and the reasons why we feel that a new algebraic characterisation for it was needed at all. Let us begin with the latter.


In keeping with the spirit of noncommutative geometry, the spectral triple Dirac operator was defined without reference to a tangent bundle and so is a purely algebraic object.  However, as explained earlier the space of all possible Dirac operators subject to all the real spectral triple conditions is too large for the noncommutative standard model to be a truly faithful extension of general relativity on the product of space-time with the internal charge-chirality space (\cite{smv}). This project suggests that the reason for this is that the internal space Dirac operator is from first principles an extended or integrated object and therefore its description is more suited to a quantum theory rather than a classical theory to begin with. So from this point of view a noncommutative description of a space-time takes it directly to the quantum theory afterall, missing out the classical general relativity on the continuous space-time as a preliminary step.  Moreover, the Fell bundle categorification allows us 
to take seriously the implication from the noncommutative standard model that to study the fermion masses is to study Dirac operators on noncommutative manifolds.

The noncommutative standard model asks for a precise mathematical characterisation for the geometrical meaning of the Higgs field, which so far goes something like this: ``a kind of a parallel transport between the two chiralities on a fuzzy bundle'' and it is thought of as a gauge field or a generalised connection on the internal space. Therefore in a quantum gravity theory on the product space, the Higgs will be the internal space component of the gravitational connection or holonomy. In the Fell bundle setting, one can understand the Higgs as a parallel transport on the real spectral triple in terms of the bundle nature of the Riemannian spin manifold encoded in the Fell bundle geometry.  It is this identity for the Higgs together with the interpretation of $D$ as a field of morphisms in a category that makes it possible to make the link with spin foam models. Moreover, if the Higgs really is a component of an enveloping particle with the graviton, a more precise meaning of what it is may be needed to 
help detect it.

\subsubsection{How $D_{FB}$ was defined}

To arrive at an intuition for a new algebraic interpretation of the Dirac operator in noncommutative geometry the following data was considered to provide clues for what it ought to be before a formal definition was written down. To check its validity we verified the classical limit in \cite{sc}. Below we also give an alternative more transparent version of the definition.

The first clue is that $D$ is a derivation of the spectral triple algebra and \emph{hence it should be an element of a module over it}. An inner derivation on a Banach algebra $A$ (\cite{Harti}) is a derivation $\delta$ from $A$ into an $A$-bimodule $N$ for $x \in N$,

\begin{equation}
 a \mapsto [x,a] ~\forall a \in A
\end{equation}

In the context of a real spectral triple, the definition of a derivation is

\begin{equation}
 \delta(ab^{opp}) = a \delta (b^{opp}) +  \delta( a)b^{opp}
\end{equation}

where $a \in A$ and $b^{opp}$ is in the opposite algebra of $A$. One of the axioms for a real spectral triple is the first order condition (see \cite{gravity} for all the axioms) and this is equivalent to the statements that $D$ defines an inner derivation and $[a, b^{opp}]=0$.

Let $E$ be a Fell bundle. $E$ is a bimodule over $E^0$ and in particular, Fell bundle $C^*$-categories are built from imprimitivity bimodules. Let us consider a finite spectral triple with algebra $A$. Note that if $a \in A = E_{\gamma \gamma^{\ast}} \oplus E_{\gamma^{\ast} \gamma}$, that is a section of the $C^*$-bundle $E^0$ and if $D$ is the self adjoint element,

\begin{equation}
\left(   \begin{array}{cc}
  0              &      X     \\
   X^{\ast}      &      0
\end{array}  \right)
\end{equation}

of \ref{linking algebra} then  $D + a$ is an element of the Morita equivalence linking algebra \ref{linking algebra}. As an alternative notation put $X = D_{FB}^+$, $X^* = D_{FB}^-$.

An elementary fact of differential geometry is that the set of sections of a tangent bundle $TM$ form a module over the algebra of functions $C(M)$ on the manifold $M$ and the tangent vector fields, the elements of $TM$, are derivations of $C(M)$. The second clue is the definition of a flat Dirac operator $D_{\mathrm{RSM}}$ on a Riemannian spin manifold is a tangent vector field contracted with the Clifford algebra. In local coordinates this can be written,

\begin{equation} \label{Berline D}
 D_{\mathrm{RSM}} = \sum_i c_i \frac{\partial}{\partial x^i}
\end{equation}

with $c_i$ indicating the Clifford algebra. A test for the relevance of the definition of Fell bundle Dirac operator $D$ is if in the classical limit as $M \times M$ goes to $TM$, $D$ takes the form given by  $D_{\mathrm{RSM}}$.

Recall that a tangent vector field is a continuous assignment of a tangent vector to each point in the manifold. An alternative definition is as a section of the domain map $d$ of $TM$ viewed as a groupoid over $M$,

\begin{equation}
 d: TM \rightarrow M
\end{equation}

Hence a deformed tangent vector field is a continuous field of groupoid arrows, or a section of the domain map $d$ of $\Gamma$,

\begin{equation}
 d: \Gamma \rightarrow \Gamma_0
\end{equation}

where the unit space $\Gamma_0$ of $\Gamma$ is of course identified with $M$. (To give a section of $d$ just means that to each object of the category one assigns one morphism that is sourced there.)

The new step we take is to replace the space of objects $\Gamma_0$ of the category $\Gamma$ with the space of objects $E^0$ (slight abuse of notation) of the category $E$. In this way we replace points with objects in a category other than Set. The objects of $E$ have a Grothendieck type topology, which we made precise in \cite{sc}. Using this picture we give the notion of ``fuzzy point'' in a noncommutative space an actual meaning as a fibre in a Fell bundle; instead of dealing with a point, one works with an object in a Fell bundle, the algebra $\mathbb{C}$ which one generalises to other fibres $M_n(\mathbb{C})$.

We define a Fell bundle Dirac operator $D$ to be a self-adjoint section of the map,

\begin{equation}
 d: E \rightarrow E^0
\end{equation}

In the special case that $C^*(E^0)$ is commutative and $M$ is a smooth Riemannian spin manifold as opposed to a noncommutative one, the Fell bundle Dirac operator $D$ is a self-adjoint section of

\begin{equation}
 d: Cl \rightarrow Cl^0
\end{equation}

modulo conditions related to real spectral triples. $Cl^0$ denotes the Clifford bundle. Note that the definition of Fell Dirac operator given in \cite{sc} was more complex because it involved the conditions necesssary to build a real spectral triple from a Fell bundle. The current purpose is to give the above more transparent definition for the more general case in order to elucidate its meaning.

In the case that $H$ is finite dimensional, $(C^*(E^0), H, D)$ is a spectral triple and in the case that $H$ is infinite dimensional and $M$ is a commutative manifold, a commutative spectral triple should be its classical limit (that is, as $\Gamma \rightarrow TM$ while $\hbar \rightarrow 0$). One of the theme of \cite{sc} was to try to prove that Commutative real spectral triples with complex algebra pertaining to a compact and simply connected Riemannian spin manifold $M$ (with
 Clifford algebra according the scope given in \cite{sc})  correspond to the classical limit of some
Fell bundle geometry.

More specifically, a finite real even spectral triple is a finite spectral triple together with two more structures $(A, H, D, J, \chi)$: A real structure on a spectral triple \cite{reality} is given by an antiunitary operator $J$ on $H$ such that $J^2 = \pm 1$, $DJ= \pm JD$, $[a,b^{opp}]=0$,
$[[D,a],b^{opp}] = 0$ where $b^{opp} = Jb^*J^*$ for all $b \in A$ is an element of $A^{opp}$. A further condition on the Dirac operator for a real \emph{even} spectral triple comes from the orientability axiom: $D \chi = - \chi D$ where $\chi$ is the $\mathbb{Z}/2$-grading on the Hilbert space and is also gives the volume form. In the Euclidean signature $\chi = (1, -1, -1, 1)$ and in the Lorentzian, $\chi = (1, -1, 1, -1)$.

As will become explicitly clear in the examples and calculations section below, for a Fell bundle geometry to characterise a real spectral triple, it needs a real structure and as a result must take the form of a product bundle $E \otimes E^{opp}$ over a groupoid $G$ where the opposite Fell bundle $E^{opp}$ is obtained by replacing the fibre over $g$ with that over $g^*$ for all $g \in G$. This is the generalisation of the product bundle $TM \otimes T^*M$.

\begin{Fell bundle geometry}
 A noncommutative  \emph{Fell bundle geometry} $(E,D)$ is a finite dimensional tensor product Fell bundle $C^*$-category $E \otimes E^{opp}$ together with a finite dimensional Hilbert space carrying a representation of the algebras of sections of $E^0$ and $E$ and  with a $\mathbb{Z}/2$-grading and reality structure and a Fell bundle Dirac operator.
\end{Fell bundle geometry}

\begin{D}
 A \emph{Fell bundle Dirac operator} $D_{FB}$ or $D$ assigned to a noncommutative finite dimensional Fell bundle geometry is a section of the domain map of a Fell bundle $C^*$-category of the form $E \otimes E^{opp}$ satisfying $D = D^* = JD^*J^{-1}$.
\end{D}

We say that a real spectral triple has a \emph{Fell bundle categorification} if it can be recast in the form of a Fell bundle geometry.

\subsubsection{Gravitational observables}

In spectral geometry, the gravitational field fluctuations are encoded in the Dirac operator, that is to say the degrees of freedom of $D$ provide the field configuration for the theory and therefore the observables arise from it are to generate a $C^*$-algebra. Therefore we pass to bounded operators (see formula \ref{exponentiated} below) as in loop quantum gravity in which the gravitational field variable (connection) is exponentiated or integrated to generate a $C^*$-algebra. The reasons in loop quantum gravity for passing to a bounded alternative (the set of holonomies separate the space of connections) is because since the space of connections $\mathcal{A}$ can be infinite dimensional, it becomes difficult to define the set of all square integral functions on it. If the basic field to be quantised is the holonomy instead, then the states can be defined in terms of paths in spin networks and the algebra of observables can be completed in the $C^*$-norm allowing the use of $C^*$-algebraic tools. Recall 
that true physical observables are self-adjoint unbounded operators on the physical Hilbert space and the truly measurable quantities are their expectation eigenvalues. Observables on an infinite dimensional Hilbert space are unbounded because they define inner derivations: this is because the relation $ab-ba=1$ can only be satisfied with $a$ and $b$  operators on an infinite dimensional Hilbert space if the operators are unbounded \cite{book}. An algebra of observables $C^*(T^*M)$ is of course bounded. To solve this, one may choose to restrict the domain of $H$ to the intersection of the domains of definition of all the unbounded operators, or exponentiate the unbounded operator to obtain a unitary. Obviously when $H$ is finite dimensional, no such circumvention is necessary.

\subsubsection{Parallel transport between the two chiralities}

With the Dirac operator of a finite triple being a matrix, the Higgs as a generalised connection is analogous to a parallel transport. The following needs to be made more precise but the Fell bundle generalisation of a path lifting in a vector bundle is a $*$-functor $pl: G \rightarrow E$. For example, it is a fact from Lie groupoid theory that a representation of a pair groupoid in $E^0$ as a vector bundle is equivalent to a parallel transport corresponding to a flat connection. In general, Fell bundles do not have isomorphic fibres and so instead of to an isomorphism, a path lifting will pertain to a homomorphism of fibres. A concrete Dirac operator (of which an example is the Higgs $\Phi \in \mathbb{C}^2$) is a representation of a Fell bundle path lifting on a Hilbert space. The Dirac operators appearing in the section on calculations in physics below are concrete. In quantum theories the Dirac operator famously maps fermionic states between the two chirality eigenspaces $H_L$ and $H_R$, or is `a map from 
left to right'. A Fell bundle Dirac operator can be identified with a homomorphism $D_{FB}^+$ from $E(L)$ to $E(R)$ (that is from the object of $E$ over the groupoid unit $L$ to that over the unit $R$). For example let $E(L) = M_2(\mathbb{C})$ and $E(R) = \mathbb{C}$, then $D_{FB}^+$ is an element of the $M_2(\mathbb{C}) - \mathbb{C}$ imprimitivity bimodule $\mathbb{C}^2$. Representing the category $E$ on the $C^*$-category Hilb including $H_L$ and $H_R$ as Hilbert space objects (for how to do this see \cite{GLR}), we attain the map from the left-handed fermions to the right-handed fermions.

Since they give rise to observables all Fell bundle Dirac operators are defined to be bounded even if the Hilbert space is infinite dimensional. Schematically, what is meant by a Fell bundle Dirac operator is a geodesic flow,

\begin{equation}    \label{exponentiated}
 D_{FB} = e^{it \vert  D  \vert}
\end{equation}

where $D$ is the usual unbounded Dirac operator probing a commutative manifold $M$, with $t \in M \times M$.

\subsubsection{Why groupoids?}

Even though groupoids do not usually appear in quantum gravity, and neither are they used in the noncommutative standard model and they do not occur in AQFTs, they are an integral part of the constructions involved in the current context. Here we explain why.

Firstly as is now well known, groupoids can provide a ``spectrum'' for noncommutative algebras; if we view a groupoid $C^*$-algebra as an algebra of observables then an arrow should be thought of as a generalised momentum. In the tangent groupoid quantisation the tangent bundle $TM$ is deformed to the pair groupoid $\Gamma = M \times M$. Fell bundles over groupoids provide a partnership between a context in which to study Morita equivalance and the tangent groupoid quantisation applied to noncommutative space-times. On discretising the quantum gravity path integral the connections on a principle bundle are replaced with holonomies or parallel transports, and these are nothing other than (fibre) isomorphisms, that is, groupoid arrows and a Fell bundle geometry over a groupoid is supposed to provide a context for a more precise definition of the Higgs parallel transport. Groupoids allow us to replace space-time with a category and to consider parallel transports as functors. Fell bundles over groupoids also 
provide a context for a definition of an orientable noncommutative bundle with non-isomorphic fibres. A category of Morita equivalence bimodules can be thought of as a ``weak'' generalisation of a groupoid\				.  A consequence to the spectral action is that if $D$ is bounded then the cut-off $\Lambda$ is not needed, that is, the scale of noncommutativity doesn't have to be arbitrarily determined by $\Lambda$.

\subsection{Calculations}

Here we work through the Fell bundle categorifications of two examples of real even finite spectral triples and consider the implications to the form of the fermion mass matrix. The specific calculations are model dependent but we expect the methods and techniques to be adaptable as the currently most up-to-date model continually evolves.



\subsubsection{Example 1}

This is a toy model for quarks without colour charge and a warm-up exercise.

The sector of the Hilbert space for the first generation of quarks, up and down has a basis denoted:

\begin{equation}
\Psi=(u_L, d_L, u_R, d_R, u_{\bar{L}}, d_{\bar{L}}, u_{\bar{R}}, d_{\bar{R}})^T
\end{equation}

Let the spectral triple algebra  be given by

\begin{equation}
  A \oplus B \oplus \alpha 1_2  \oplus \beta 1_2  ~~  \in   M_2(\mathbb{C})  \oplus  M_2(\mathbb{C}) \oplus \mathbb{C}  \oplus \mathbb{C}
\end{equation}

The opposite algebra is defined using the reality structure,

\begin{equation}
  \lambda 1_2 \oplus \mu 1_2 \oplus E  \oplus F ~~  \in  \mathbb{C}  \oplus \mathbb{C}  \oplus   M_2(\mathbb{C})  \oplus  M_2(\mathbb{C})
\end{equation}

To proceed with the categorification of the spectral triple, let $(E \otimes E^{opp}, \pi, \Gamma)$ be a Fell bundle with where $\Gamma$ the pair groupoid on the discrete space consisting of the 4 points $\Gamma$ = Pair(4) and let the spectral triple algebra be identified with $C^*(E^0)$. A section of $E^{0} \otimes E^{opp,0}$ is identified with an element of a representation of $C^*(E^{0} \otimes E^{opp,0})$ on the Hilbert space,

\begin{equation}
 a \otimes b^{opp} = A \otimes \lambda  ~~ \oplus ~~   B \otimes \mu   ~~  \oplus  ~~  \alpha \otimes E  ~~ \oplus ~~ \beta \otimes F ~~ \in A \otimes A^{opp}
\end{equation}



The sum of the element $a \otimes b^{opp}$ and a choice of a self-adjoint section of the domain map $d : E \otimes E^{opp} \rightarrow  E^0 \otimes E^{opp,0}$ gives a section of $E \otimes E^{opp}$,

\begin{equation}     \label{no particle exchange}
\left(   \begin{array}{cccc}
  A \otimes \lambda               &      X  \otimes c          &        0                     &                 0                  \\
   X^{\ast} \otimes \bar{c}       &      B \otimes \mu         &        0                     &                 0                  \\
 0                                &              0             &  \alpha     \otimes F        &           a  \otimes Y           \\
 0                                &              0             &   \bar{a}      \otimes Y^*   &          \beta  \otimes G
\end{array}  \right)
\end{equation}

with $X,Y \in M_2(\mathbb{C})$, $a,c \in \mathbb{C}$.

Then if we apply the condition that $DJ = JD$,

\begin{equation}
D = \left(   \begin{array}{cccc}
          0                          &       X  \otimes c      &        0                &                 0                  \\
  X^{\ast} \otimes \bar{c}        &              0             &        0                &                 0                  \\
 0                                   &              0             &        0                &           \bar{X} \otimes \bar{c}            \\
 0                                   &              0             &   X^T \otimes c         &                 0
\end{array}  \right)
\end{equation}

which implies the condition that:

\begin{equation}
 a  \otimes Y =  \bar{X} \otimes \bar{c} = M ~  \in M_2(\mathbb{C}),
\end{equation}

and so this Dirac mass matrix $M$ (map between left and right) is an element of $M_2(\mathbb{C})$.

At this point it has become clear why the reality structure on the Fell bundle was needed, that is, why a Fell bundle geometry must be a product bundle to satisfy the reality condition on $D$.

\subsubsection{Example 2}

In \cite{sap}, \cite{reality} a faithful representation of the algebra below is given to precisely capture all the fermion charges. Alternatives to this algebra also appear in the literature provided they can be represented to give also the right charges to reflect those seen in physics.

\begin{equation}
 A = \mathbb{H}  \oplus   \mathbb{C}  \oplus  M_3(\mathbb{C})
\end{equation}

In some models this is extended to four or more summands,

\begin{equation}
 A =  \mathbb{H}  \oplus \mathbb{H} \oplus   \mathbb{C}  \oplus  M_3(\mathbb{C})
\end{equation}

or since $M_2(\mathbb{C})$ is isomorphic to $\mathbb{H}  \oplus \mathbb{H}$,   $A$ may be given as

\begin{equation} \label{algebra}
 A = M_2(\mathbb{C})  \oplus   \mathbb{C}  \oplus  M_3(\mathbb{C}).
\end{equation}

The final answer for the algebra is not known yet, and Connes hints that it will involve a q-deforming in \cite{gravity}.

In this calculation we will consider a faithful representation of \ref{algebra} because it is over the complex numbers, otherwise we would have to work with real Fell bundles.  $A$ is not necessarily identified with $C^*(E^0)$ but the latter should be represented to give the same charges if not faithfully.

Now we give the specifications for this example of a real even finite spectral triple $(A, H, D, J, \chi)$.

Restricting to one family or the first generation fermions, that is, to ``normal matter'' leaving the full set of 3 generations of fermions for possible further work. We treat the quarks first and then add leptonic matter.

The basis of the Hilbert space is given by the single generation quark sector of the standard model Hilbert space $\mathbb{C}^{24}$,

\begin{displaymath}
 \Psi= (\Psi_L, \Psi_R, \Psi_{\bar{L}}, \Psi_{\bar{R}})^T =
\end{displaymath}

\begin{equation}
(u^r_L, d^r_L, u^g_L, d^g_L, u^b_L, d^b_L, u^r_R, d^r_R, u^g_R, d^g_R, u^b_R,   d^b_R, u^r_{\bar{L}}, u^g_{\bar{L}}, u^b_{\bar{L}}, d^r_{\bar{L}}, d^g_{\bar{L}}, d^b_{\bar{L}}, u^r_{\bar{R}}, u^g_{\bar{R}}, u^b_{\bar{R}}, d^r_{\bar{R}}, d^g_{\bar{R}}, d^b_{\bar{R}})^T
\end{equation}

with $\mathbb{Z}/2$-grading $\chi$ and reality $J$. The Dirac operator is the mass matrix of the one generation quark sector from standard model physics and consists of blocks of 6 by 6 matrices including $M$, which is a map from $\Psi_R$ to $\Psi_L$. The matrix $M$ is a 6 by 6 matrix, so it has 36 free parameters but as quarks that only differ by colour charge have the same mass, there are only two different masses plus the Higgs doublet, making only 4 degrees of freedom required.

These are the faithful representations of the spectral triple algebra on the Hilbert space

\begin{equation}
  A 1_3  \oplus \bar{\lambda} 1_3 \oplus \lambda 1_3 \oplus n 1_2 \oplus n 1_2    \in A
\end{equation}

\begin{equation}
  m 1_2 \oplus m  1_2  \oplus    F 1_3   \oplus  \bar{\mu} 1_3    \oplus    \mu 1_3    \in A^{opp}
\end{equation}

And the tensor product algebra acts faithfully on the Hilbert space as $(a \otimes b^{opp}) \Psi = a \Psi b^{opp}$, with $b^{opp} = J b^* J^{-1} \in A^{opp}$ for all $b \in A$,

\begin{equation}
  A  \otimes  m ~~  \oplus   ~~    \bar{\lambda}   \otimes m   ~~\oplus ~~    \lambda \otimes m ~~ \oplus ~~ n  \otimes F   ~~  \oplus ~~  n  \otimes \bar{\mu}    ~~  \oplus ~~  n \otimes \mu
\end{equation}

note that we omitted the $1_3$ and $1_2$ to ensure that each block is a 6 by 6 matrix. And this defines for us the algebra $C^*(E^0 \otimes E^{0, opp})$.



The sum of the element $a \otimes b^{opp}$ and a Fell bundle Dirac operator giving a section of $E \otimes E^{opp}$:

\begin{equation}
\left(   \begin{array}{cccc}
  A  \otimes m                      &      X   \otimes p         &        0                     &                 0                  \\
   X^{\ast}  \otimes \bar{p}        &  \Lambda  \otimes m        &        0                     &                 0                  \\
 0                                  &              0             &     n \otimes F              &             l  \otimes Y           \\
 0                                  &              0             &   \bar{l}     \otimes Y^*    &           n \otimes \Lambda'
\end{array}  \right)
\end{equation}

$\Lambda$ and $\Lambda'$ are faithful embeddings of $\mathbb{C}$ into $M_2(\mathbb{C})$,

\begin{equation}
 \Lambda,~ \Lambda' : \mathbb{C}  \rightarrow M_2(\mathbb{C})
\end{equation}

\begin{equation}
 \lambda \mapsto    \left(   \begin{array}{cc}
  \bar{\lambda}       &       0     \\
            0         &       \lambda
\end{array}  \right), \qquad
\mu \mapsto    \left(   \begin{array}{cc}
  \bar{\mu}       &       0     \\
            0         &       \mu
\end{array}  \right)
\end{equation}

and $X$ and $Y$ are any faithful embeddings of $\mathbb{C}^2$ into $M_2(\mathbb{C})$.

\begin{allowed D}[Absence of leptoquarks]
The sections $x$ of the domain map $d$ of a Fell bundle $C^*$-category over the groupoid $Pair(4)$ satisfying $x=x^*$ and $x= J x^* J ^{-1}$ can take any of and only the following forms:
\end{allowed D}

\begin{equation}     \label{canonical choice}
\left(   \begin{array}{cccc}
0         &          M          &      0        &   0        \\
M^*       &          0          &      0        &   0        \\
0         &          0          &      0        &  \bar{M}   \\
0         &          0          &      M^T      &  0
\end{array}  \right) , \qquad
\left(   \begin{array}{cccc}
0         &          0          &   K        &   0\\
0         &          0          &   0        &   H\\
K^*       &          0          &   0        &  0\\
0         &          H^*        &   0        &  0
\end{array}  \right),
\end{equation}

\begin{equation}  \label{w}
\left(   \begin{array}{cccc}
A         &          0          &   0        &   0\\
0         &    \Lambda          &   0        &   0\\
0         &          0          &   \bar{A}  &  0\\
0         &          0          &   0        &  \bar{\Lambda}
\end{array}  \right), \qquad
\left(   \begin{array}{cccc}
0         &          0          &   0        &     G \\
0         &          0          &   G^T        &   0\\
0         &          \bar{G}    &   0        &  0\\
G^*   &          0          &   0        &  0
\end{array}  \right)
\end{equation}

where in this specific example $M$, $G$, $K$ and $H$ are 6 by 6 matrices of which $K$ and $H$ are symmetric. The proof of the statement is obtained simply by writing out the matrices. (An effectively identical calculation was demonstrated with diagrams in the ``2-point space'' example detailed in \cite{sc}.) Of the above available set we choose the first to build the Fell bundle geometry because it provides non-zero Dirac masses: considering the Hilbert space  basis, $M^*$ we find that is a map from $\psi_L$ to $\psi_R$.

Notice that in a section of the domain map of a finite Fell bundle $C^*$-category there appears only one block in each row and column of blocks whereas the general Dirac operator in the noncommutative standard model takes the form:

\begin{equation}   \label{Euc}
D = \left(   \begin{array}{cccc}
0         &          M          &   0        &     G \\
M^*       &          0          &   G^T      &   0\\
0         &          \bar{G}    &   0        &  \bar{M}\\
G^*       &          0          &   M^T      &  0
\end{array}  \right)
\end{equation}

in the Euclidean signature and

\begin{equation}   \label{Ltz}
D = \left(   \begin{array}{cccc}
0         &          M          &   K        &   0\\
M^*       &          0          &   0        &   H\\
K^*       &          0          &   0        &  \bar{M}\\
0         &          H^*        &   M^T      &  0
\end{array}  \right)
\end{equation}

in the Lorentzian signature, with $H = H^T$ and $K = K^T$. (In the one generation quark sector, the matrices \ref{Euc} and \ref{Ltz} are 24  by 24 matrices modulo the conditions $D = D^*$, $DJ=JD$, $D \chi = - \chi D$. \cite{smv}, \cite{lng}.)

The point is that \ref{Euc} and \ref{Ltz}  are \emph{not} Fell bundle Dirac operators because they are not sections of the domain map $d$. This means we have the result that due to the current framework the unphysical leptoquarks (\cite{leptoquarks}) $G$, $K$, $H$ are excluded by construction of the Dirac operator. (The presence of the leptoquarks in the action of a physical theory is unacceptable because they effect spontaneous colour symmetry breaking which results for example in the photon gaining a mass.)

In the following step it becomes clear why the product bundle and the real structure were needed in the definition of Fell bundle geometry to satisfy the reality condition on $D$: Now apply that condition $DJ=JD$,

\begin{equation}
   l  \otimes Y    =       \bar{X}  \otimes \bar{p}        \in M_6(\mathbb{C})
\end{equation}

So on one side we have a 3 by 3 matrix tensored with a 2 by 1 matrix (two rows and one column) (represented by a 2 by 2 matrix $Y$) while on the other side of the equation we have a 2 by 1 matrix (embedded in $M_2(\mathbb{C})$ via $X$) tensored with a 3 by 3 matrix. This reduces the number of complex parameters in $M$ from 36 (for a general 6 by  matrix over $\mathbb{C}$) down to $9 + 2 = 11$.










Now to include the leptonic sector.

The diagram illustrates the space of object fibres $E^0$ of the Fell bundle $C^*$-category $E$ for the quark and lepton sectors together,

\vspace{0.5cm}

\begin{figure}[ht]   \label{d}
\xymatrix{
 M_2(\mathbb{C}) \ar@{-}[d] \ar@{}[r]  &    \mathbb{C} \ar@{-}[d] \ar@{}[r] &   M_3(\mathbb{C}) \ar@{}[r] \ar@{-}[d]  &    M_3(\mathbb{C}) \ar@{-}[d]   &  M_2(\mathbb{C}) \ar@{-}[d] &  \mathbb{C} \ar@{-}[d]  &   \mathbb{C} \ar@{-}[d]  &     \mathbb{C} \ar@{-}[d]   \\
q_L  \ar@{}[r] &  q_R \ar@{}[r] &  q_{\bar{L}} \ar@{}[r]   &    q_{\bar{R}}   &  l_L  \ar@{}[r] &  l_R \ar@{}[r] &  l_{\bar{L}} \ar@{}[r]   &    l_{\bar{R}}   }

\caption{The space of objects $E^0$}
\end{figure}
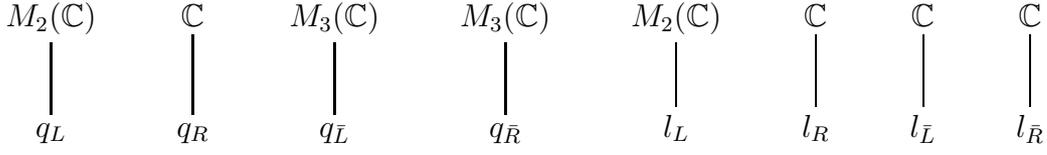

\vspace{0.5cm}

Now the abstract algebra $C^*(E^0)$ of the $C^*$-bundle $C^*(E^0)$ is not equal to $A$, but we can still represent it in the same way on the Hilbert space although now it will not be a faithful representation.

\begin{equation}
 C^*(E^0) = M_2(\mathbb{C}) \oplus  \mathbb{C} \oplus M_3(\mathbb{C}) \oplus  M_3(\mathbb{C}) \oplus M_2(\mathbb{C}) \oplus  \mathbb{C} \oplus  \mathbb{C} \oplus  \mathbb{C}
\end{equation}

The basis of the Hilbert space for the first generation lepton sector is

\begin{equation}
 \Psi = (\Psi_L, \Psi_R, \Psi_{\bar{L}}, \Psi_{\bar{R}}) = (e_L, \nu_L, e_R,  e_{\bar{L}}, \nu_{\bar{L}}, e_{\bar{R}} )  ~~ \in \mathbb{C}^6
\end{equation}

The algebra representation for the first generation of leptons:

\begin{equation}
 a = A    \oplus    \bar{\lambda}   \oplus   0   \oplus   \lambda   \oplus   \lambda ~~ \in M_2(\mathbb{C}) \oplus \mathbb{C}
\end{equation}

\begin{equation}
 b^{opp} = \mu  \oplus  \mu   \oplus  B   \oplus   \bar{\mu}  \oplus 0  ~~   \in \mathbb{C}   \oplus  M_2(\mathbb{C})
\end{equation}

The zeroes are only included to keep the matrices nice and square. Simply removing the rows and columns of zeroes and omitting the $\nu_R$ from the Hilbert space ensures that the Poincar\'e duality axiom of real spectral triples is obeyed.

Again we assign to the Fell bundle the Dirac operator $D$ that provides both the quark and lepton sectors with non-zero Dirac masses (maps between left and right). Since only one block may appear in each row and column of blocks, we find that there are no non-zero maps $m$ to allow mixing between quarks and leptons $m: q \rightarrow l$.

One section $D + a \otimes b^{opp}$ of the tensor product Fell bundle restricted to the lepton sector is:-

\begin{equation}
\left(   \begin{array}{cccc}
 A  \otimes \mu                     &      W   \otimes f                 &        0                     &                 0                  \\
 W^{\ast}  \otimes \bar{f}          &              r                     &        0                     &                 0                  \\
 0                                  &              0                     &     \lambda \otimes B        &             g  \otimes Z           \\
 0                                  &              0                     &   \bar{g}     \otimes Z^*    &                 s
\end{array}  \right)
\end{equation}

where

\begin{displaymath}
   r = \left(  \begin{array}{cc}
      \bar{\lambda}  \otimes \mu    &   0   \\
          0                         &   0
\end{array} \right), \qquad
   s= \left(  \begin{array}{cc}
    \lambda \otimes \bar{\mu}       &   0    \\
          0                         &   0
              \end{array}   \right), \qquad  W, Z \in \mathbb{C}^2, \qquad f,g \in \mathbb{C}.
             \end{displaymath}

Imposing the condition $DJ=JD$, we have

\begin{equation}
M =  g \otimes Z = \bar{W} \otimes \bar{f}  \in \mathbb{C}^2
\end{equation}

which gives the correct number of degrees of freedom for the Higgs doublet giving one mass to the electron and none to the neutrino (one sees this by diagonalising $M$ to put the Hilbert space into mass eigenstates).

\subsubsection{Conclusions}    \label{conclusions}

\begin{enumerate}
 \item The first generation part of the standard model finite spectral triple (at least in the model we used) has a Fell bundle geometry categorification, and in which:
     \begin{enumerate}
           \item   the vestige of unphysical leptoquarks is absent,
           \item   as in physics, quark-lepton mixing $m: q \rightarrow l$ does not take place,
           \item   the number of extra free parameters in the mass matrix $M$ is reduced.
          \end{enumerate}
\item The result that $W, X \in \mathbb{C}^2$  relates the fact that $\mathbb{C}^2$ is a Morita equivalence bimodule over $M_2(\mathbb{C})$ and $\mathbb{C}$ with the fact that the Higgs  $( \phi_1, \phi_2 )^T \in \mathbb{C}^2$.
\end{enumerate}

\subsection{A discussion on quantisation for spectral gravity}

Below we try to discuss some topics towards quantising spectral triple gravity (naively) using ideas from non-perturbative quantum gravity and algebraic quantum field theory as our guiding principles. Recall that spectral gravity is already background free and generally covariant. Instead of the metric or even the connection playing the role of the gravitational field, in spectral geometry it is the degrees of freedom of the Dirac operator.  In this ``toy'' quantisation scheme we replace $D$ with the Fell bundle Dirac operator in analogy with replacing the connection with its holonomy while since the Dirac operator encodes both the connection and the driebeins, the space of Dirac operators also generates the full algebra of the conjugate variables (the analogy of the algebra of $p$s and $q$s). The quantisation is provided by the tangent groupoid quantisation (for the classical limit calculation see \cite{sc}). More specifically, the Fell bundle Dirac operator plays a role similar to that of a parallel 
transport and its square plays the role of the holonomy of the connection. Many of the mathematical details still need to be made more precise which is why the title is given as an outlook. We also try to begin to answer some of the questions raised in the introduction and in many ways we draw from results reviewed in part I.

\subsubsection{Kinematics}

We define the kinematics of a \textit{Fell bundle quantum system} over a Fell bundle $(p, E, \Gamma)$ to consist of algebras $C^*(E^0)$ or $A$, and $C^*(E)$, a Hilbert space $H$ to carry a representation of those two algebras, arising for example from the GNS construction for $A$, together with a space of Fell bundle Dirac operators $\{D_{FB}\}$ such that in the case that $H$ is finite dimensional, $(A,H,D_{FB})$ is a spectral triple. (If $H$ is infinite dimensional then $D_{FB}$ has the interpretation as the exponentiated Dirac operator or geodesic flow of a spectral triple.)

Recall that the observable algebra in the spectral triple loop quantum gravity formulation (reviewed above in section 2.4) is given as the $C^*$-completion of the algebra generated by a pre-defined set of holonomies, and even though in that context the Dirac operator probes the space of connections rather than the space-time, for our situation we observe a similar fact,

\begin{generating set} The algebras $C^*(E)$ and $C^*(E^0)$ are respectively the $C^*$-completions of the algebra generated by set of Fell bundle Dirac operators, and that generated by their squares.
\end{generating set}

To see the \textit{proof} of the above statement about $C^*(E)$, simply do finite sums of the form $\sum c_i D_{FB,i}$ (the $c_i$s complex numbers) with the $D_{FB,i}$s from the set of sections of the Fell bundle that satisfy the definition of Fell bundle Dirac operator and that for $C^*(E^0)$ comes from the definition of a Fell bundle, in the fact that $e_1 e_2 \in E_{12}$, in particular, $e_{\gamma} e_{\gamma^*} \in E_{\gamma \gamma^*}$. One can also see this by squaring the matrices in the set \ref{canonical choice}, \ref{w}.

So the algebra $C^*(E^0)$ is generated by the ``holonomy loops'' (locally $e e^* \oplus e^* e \in E^0$) defined by $D^2$ for all sections of $E$ satisfying the definition of a Dirac operator on a Fell bundle and $C^*(E)$ (the ``algebra of $p$s and $q$s'') is generated by that space of all possible $D_{FB}$. Recall that $C^*(E^0)$ is our noncommutative generalisation of $C^*(M)$, the $C^*$-completion of the functions on the manifold and note that when $M$ is a true manifold (as opposed to a noncommutative one) $C^*(E^0)$ is commutative and therefore the operators are multiplication operators like the algebra of $q$s in  quantum mechanics. With these interpretations of the Dirac operator, and the algebras $C^*(E)$ and $C^*(E^0)$ we can borrow the result from Aastrup, Grimstrup and Nest (see 2.4 above) that the interaction between $D_{FB}$ and $C^*(E^0)$ reproduces the Poisson structure of general relativity.

\subsubsection{Dynamics}

Secondly, the dynamics consists of expectation values for the Dirac operator spectrum, which means evaluation of linear functionals on $A$,

\begin{equation}
 \omega_{\rho}(D_{FB}^2)  =  \mathrm{Tr} \rho D_{FB}^2
\end{equation}

with $D_{FB}^2 \in A$ for $\rho$ a projection onto a pure state. This means that we can begin to understand a mathematical description of the physical intuition that the fermion masses are the fundamental excitations - or quanta of curvature - of the internal space component of the gravitational field.  In order to calculate these expectation values we will need to discretise Connes's spectral path integral. Obviously we have not solved this but the first step at least for this context is to replace $D^2$ with the square of the Fell bundle Dirac operator, which is analogous to the passing from the connection as the configuration variable for the gravitational field to its holonomy, which is how path integrals are discretised in the path integral formulation of quantum gravity.

Restricting \ref{spectral path integral} to the action $S= \mathrm{Tr}(f(D))$,

\begin{equation}   \label{restricted}
 Z = N \int e^{-\mathrm{Tr}(f(D))} D[D]
\end{equation}

The first step in discretising \ref{restricted} is to replace $D$ for the spectral triple as the gravitational field variable with $D_{FB}$, which effectively means

\begin{equation}
 D \mapsto e^{i \gamma \vert D \vert} \qquad \gamma \in \Gamma
\end{equation}

where as usual $\Gamma$ is a pair groupoid over a compact simply connected manifold or over a discrete space.

This is the analogous step to integrating the connection $A$ to a parallel transport in particular a holonomy,

\begin{equation}
 A \mapsto \mathrm{hol}(A) \in G.
\end{equation}

where $G$ is the holonomy group (which can be the structure group) of a pre-defined principle  bundle.

The next step is guided by spin foam models where a product is taken over all holonomies occurring at a given base point and the integral becomes a finite sum over geometries counted by representations of the group $G$. In the current context we count the sum over geometries by the states of the algebra $C^*(E)$ using the fact from Tomita-Takesaki theory that there is a unique 1-parameter group of automorphisms $\sigma_t^{\omega}$ for each faithful state $\omega$ which we interpret as a Fell bundle Dirac operator with $t=\gamma$. The latter needs to be made more precise but the meaning is that an automorphism of $C^*(E)$ induces an isomorphism (or homomorphism if they are not isomorphic) between fibres of $E^0$ and therefore can be associated to a Fell  bundle Dirac operator. In this way we draw from the thermal time hypothesis (see review above) such that with this interpretation, the space of all $D$s is narrowed down even further.

Restricting $\Gamma$ to Pair(2) = $\{ \gamma, \gamma^*, \gamma \gamma^*, \gamma^* \gamma \}$ and with $e \in E_{\gamma}$ and $e^* \in E_{\gamma^*}$  we define

\begin{equation}
 Z = \sum_{\omega} \prod \mathrm{Tr} (e^*e \oplus ee^*)
\end{equation}

and in general for $\Gamma$ the pair groupoid over any discrete space or over any compact simply connected manifold,

\begin{equation}
 Z =   \sum_{\omega} \prod \mathrm{Tr}(D_{FB}^2)
\end{equation}

Since we are summing over them, $Z$ doesn't depend on the state and therefore the result is an algebra invariant. In space-time quantum gravity, of course $Z$ is a topological invariant in space-time quantum gravity, and here in algebraic quantum gravity one expects that $Z$ should yield an algebra invariant. Note that as a topological invariant doesn't depend on the metric, an algebra invariant doesn't depend on the state. Many similar (and more erudite) ideas were discussed by others in \cite{BCL survey 2010},\cite{Paolo}.

With the association made above, a closely related measure of the average of the spectrum of the Dirac operator over the set of states of the algebra is Connes's spectral invariant of a von Neumann algebra:

\begin{equation}
  \bigcap_{\omega} \Sigma(\sigma^{\omega}).
\end{equation}

The problem of ultraviolet divergences is not encountered in the setting of spectral geometry and there also is no infrared divergences problem because we didn't need to choose any triangulation prescription as the underlying mathematical structure of the space-time is categorical (see \cite{sc} for specific details of the Grothendieck topology we used in that first part of the current project).

\subsubsection{AQFTs and A-TQFTs}

Here is an outlook towards a background free generally covariant Algebraic quantum field theory connected with an algebraic analogue of spin foam quantum gravity.

Let the analogues of 3-geometries or spin networks be the object algebras $\bigoplus E(t)$ of $E$, that is regions in the Grothendieck topology and the analogues of the graphs in the 4-manifolds be the morphisms of $E$.  Now pass to the category of representations $\mathrm{Rep}(E)$ of the objects of $E$ (regions of the Grothendieck topology) on a concrete $C^*$-category $\mathcal{U}$ such as Hilb. The morphisms are intertwiners, which are given by mapping functorially the morphisms of $E$ to those of $\mathcal{U}$. $\mathrm{Rep}(E)$ and Hilb are tensor or monoidal $C^*$-categories. $\mathrm{Rep}(E)$ is reminiscent of the DHR categories of ``local transportable morphisms'' in AQFT, see \cite{Haag's book}.

Now we can define an algebraic TQFT or an A-TQFT to be a symmetric monoidal $*$-functor $Z: \mathrm{Rep}(E) \rightarrow \mathcal{U}$.

In this outlook, the Wightman fields are substituted by sections of a noncommutative bundle and therefore smearing is not required to take account of the fact that an observable cannot be evaluated at a point. The quantisation scheme described above proceeds in the spirit of AQFT, in which the entire theory is encoded in the algebra - the states, the Hilbert space, and even the dynamics.

\subsubsection{Dictionary of terms in non-perturbative space-time QG versus AQG}

This table was compiled by surmising the ideas described in the review section and by including some of the features of Fell bundle quantum systems.

\begin{center}
\begin{tabular}[c]{|l|l|} \hline
\textbf{Space-time QG }  &  \textbf{Algebraic QG} \\ \hline
 diffeomorphism          &   algebra automorphism \\ \hline
 connection & spectral triple Dirac operator \\ \hline
 holonomy   & Fell bundle Dirac operator squared \\  \hline
 parallel transport   &   groupoid representation  \\ \hline
 triangulated space-time & Grothendieck topology or quantum topos \\      \hline
 principal bundle &  Fell bundle  \\       \hline
 path integral    & spectral path integral \\  \hline
 TQFT             & A-TQFT \\  \hline
 topological invariant   & algebra invariant \\ \hline
 sum over group representations    & sum over $C^*$-algebra states   \\ \hline
 graviton                   &      graviton plus Higgs composite  \\ \hline
 issue of time              &      state dependent notion of time-flow \\ \hline
 loop observable algebra    &      $C^*(E^0)$ \\ \hline
 quantum algebra of conjugate variables     &      $C^*(E)$  \\ \hline
 \end{tabular}
\end{center}

\subsubsection{Fermionic matter inclusion}

We make the physical interpretation that the elements of $C^*(E)$ are observables pertaining to the force of Higgs gravity acting on the fermions as represented by a Hilbert space coming from the GNS construction for that algebra. The relationship between this Hilbert space and that of the square integrable sections of the spinor bundle needs to be clarified. In the noncommutative standard model the eigenvalues of $D$ are directly related to the work done against the Higgs force in transporting a fermion between the two chiralities in the noncommutative manifold. This is encoded by open path in the space $E$, in other words a morphism. These observables are not diffeomorphism invariant, as they shouldn't be, because the fermions live on the space experiencing the pull of gravity; they are not the intrinisic geometric degrees of freedom for gravity.

\section{Discussion}

It is an old idea that geometry is well studied using algebraic methods but to date this is not a mainstream approach of quantum gravity. Perhaps the methods of Connes where he treats algebra almost as a laboratory will be important for the future progress of quantum gravity.

From the considerations of this paper, a question that stands out is that of which is the correct choice of a von Neumann algebra from which to extract the dynamics of the theory. We are also led to ask whether given a von Neumann algebra and a state, if the Tomita-Takesaki theory can be used to reconstruct a $C^*$-bundle and a connection.

In terms of Fell bundle quantum systems, problems include to (i) build a specific model for example by choosing a von Neumann algebra and constructing the Hilbert space of geometrical states from the GNS construction for $C^*(E^0)$ and the fermionic Hilbert space from that of $C^*(E)$, ultimately to build a theory on the product of space-time with internal space, (ii) understand the relation between $Z$ and other algebra invariants, (iii) make certain features of the theory as indicated more precise and (iv) construct an AQFT for spectral gravity.

The 4d space-time projection of the classical limit of the final theory will obviously be very close to general relativity but the classical gravity may not necessarily apply to internal space at all. Moreover, internal space is not described on its own separated from space-time and it may also be true that 4d quantum gravity doesn't make sense without the interaction with internal space.

\section*{Acknowledgements}

Thanks to John Barrett, Paolo Bertozzini and Pedro Resende for discussions. 

Research supported by Funda\c{c}$\tilde{\mathrm{a}}$o   para as Ci\^encias e a Tecnologia (FCT)
including programs POCI 2010/FEDER and SFRH/BPD/32331/2006.

\end{document}